\documentclass[aip,pop,reprint,groupedaddress]{revtex4-1}
\usepackage{eurosym}
\usepackage{amssymb,amsmath}
\usepackage{graphicx}

\graphicspath{ {./figures/} }
\begin{document}

\title{Modeling of reduced secondary electron emission yield from a foam or fuzz surface}
\author{Charles Swanson and Igor D. Kaganovich \\
\textit{Princeton Plasma Physics Laboratory, Princeton University,
Princeton, New Jersey 08543, USA}}
\date{September 5, 2017}
\maketitle

\address{\textit{Princeton Plasma Physics Laboratory, Princeton University,
Princeton, New Jersey 08543, USA}}

\section{Abstract}

Complex structures on a material surface can significantly reduce the total secondary
electron emission yield from that surface. A foam or fuzz is a solid surface above which is placed 
a layer of isotropically aligned whiskers. Primary electrons that penetrate into this 
layer produce secondary electrons that become trapped and not escape into 
the bulk plasma. In this manner the secondary electron yield (SEY) may be reduced. 
We developed an analytic model and conducted numerical simulations of secondary 
electron emission from a foam to determine the extent of SEY reduction. We find that 
the relevant condition for SEY minimization is $\bar{u} \equiv AD/2 >>1 $, where $D$ 
is the volume fill fraction and $A$ is the aspect ratio of the whisker layer, the ratio of 
the thickness of the layer to the radius of the fibers. We find that foam can not reduce 
the SEY from a surface to less than 0.3 of its flat value.

\section{Introduction}

\label{sec:intro}

Secondary electron emission (SEE) from dielectric and metallic surfaces can significantly 
change the electric potential profiles and fluxes near that surface. In low-temperature 
plasma applications, SEE may limit the total throughput. Examples are RF amplifiers \cite{multipactor}, 
particle accelerators, and Hall thrusters \cite{raitses2011,Wang2014,Raitses2013}. Texturing the geometry of the walls of the 
device to reduce the secondary electron yield (SEY) is an area of active research. Examples 
of recent subjects of research are triangular grooves\cite{stupakov,wang,pivi,suetsugu}, oxides\cite{ruzic}, dendritic structures\cite{baglin}, 
micro-porous structures\cite{ye2017}, and fiberous structures.

Such fiberous structures can include velvet, feathers, and foam. Fiberous structures are layers of whiskers 
grown onto a surface. In a velvet, the whiskers are all aligned in one direction, usually normal to 
the surface\cite{aguilera,Huerta,swanson2016,Raitses2013}. In a previous 
publication\cite{swanson2016}, we predicted that velvets are well-suited to minimizing SEY from a distribution of 
primary electrons which are normally incident. In this case
the reduction factor can be $<10\%$. 

Note: In this paper ``reduced by $n\%$'' means $\gamma \rightarrow (1-\frac{n}{100\%})\gamma$ 
and ``reduction factor of $n\%$'' means $\gamma \rightarrow \frac{n}{100\%} \gamma$.

In a feathered surface, the whiskers are also aligned normally and have 
smaller whiskers grown onto their sides.  In a previous publication\cite{swanson2017}, 
we predicted that these secondary whiskers serve to reduce SEY from more shallowly incident 
primary electrons and allow a more isotropic reduction in SEY.

In foam, and closely related fuzz, the whiskers are isotropically aligned, producing a random 
layer of criss-crossing whiskers\cite{cimino,Huerta,patino2016}. The SEY from fuzz/foam is of interest to the low-temperature 
plasma modeling community because it is expected 
to behave more isotropically than the uniformly aligned fibers of velvet. The SEY from fuzz/foam 
is of interest to the high-temperature plasma modeling community because 
tungsten fuzz is spontaneously generated in the tungsten divertor region of Tokamak 
plasma confinement vessels.

Recent experiments on this self-generated tungsten fuzz reports SEY reduction factors of $40\%-60\%$ 
and little dependence on the primary angle of incidence \cite{patino2016}. 

Copper fuzz/foam was simulated using a Monte-Carlo algorithm recently \cite{Huerta}. The  
geometry used was a ``cage'' geometry consisting of normally aligned whiskers and perfectly 
regular, rectangularly placed, horizontal whiskers. Using this approximation 
and geometrical values taken from experimental characterization of real foams, the 
reduction factor was calculated to be $~70\%$.

In this paper, we report the results of numerical simulations of SEY from a foam surface. 
Furthermore we apply a simplified analytic model to explain the results. The numerical values in this 
paper will be given assuming a carbon graphite surface. However, according to the analytical model, 
the SEY reduction is not dependent on material.

\section{Numerical model}

\label{sec:numerical}

We performed a Monte Carlo calculation of the SEY of a foam surface. We
used the same simulation tool that was previously used to simulate SEY from
velvet and was benchmarked against analytical calculations \cite{swanson2016}.

We numerically simulated the emission of secondary electrons by using the
Monte Carlo method, initializing many particles and allowing them to follow
ballistic, straight-line trajectories until they interact with the surface.
The surface geometry was implemented as an isosurface, a specially designed
function of space that gives correct structure. The SEY of a
particle interacting with a flat surface was assumed to follow the empirical
model of Scholtz, \cite{scholtz} 
\begin{equation}
\gamma \left( E_{p},\theta \right) =\gamma _{max}(\theta )\times \exp \left[
-\left( \frac{\ln [E_{p}/E_{max}(\theta )]}{\sqrt{2}\sigma }\right) ^{2}%
\right] .  \label{gamma}
\end{equation}

Secondary electrons were assumed to be emitted with probability weighted 
linearly with normal velocity component (cosine-law emission) \cite{Bronstein}.

For parameters in the model $\gamma _{max}(\theta )$, $E_{p}$, $%
E_{max}(\theta )$, $\sigma ,$ we used those of graphite \cite{patino2013},
assuming structures are carbon based.  The form of the angular dependence 
$\gamma_{max}(\theta), E_{max}(\theta)$ is taken from Vaughan \cite{vaughan} 

\begin{equation}
\label{eq:angularsey}
\gamma _{max}(\theta )=\gamma _{max_{0}}\left( 1+\frac{k_{s}\theta ^{2}}{2\pi
}\right),
\end{equation}
\begin{equation*}
E_{max}(\theta )=E_{max_{0}}\left( 1+\frac{k_{s}\theta ^{2}}{\pi
}\right) .
\end{equation*}

We initialized the primary electrons
with an energy of 350eV. True secondary electrons, elastically scattered
electrons, and inelastically scattered electrons were taken into
consideration. For more discussion on the
model and its implementation in the Monte Carlo calculations, see our
previous paper on SEE from velvet \cite{swanson2016}.

\begin{figure}[tbp]
\centering\includegraphics[scale=0.45]{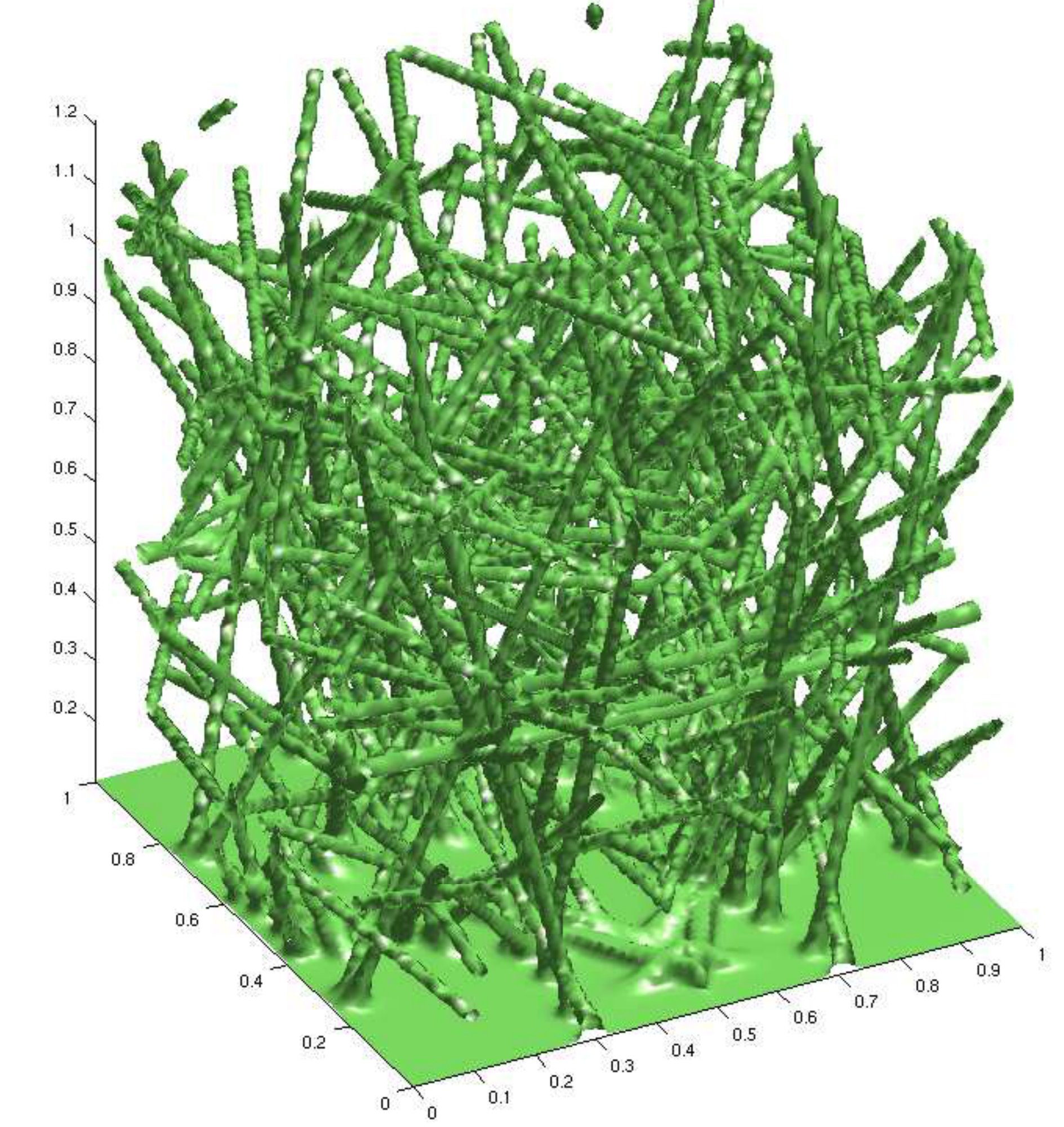}
\caption{Rendering of an example of the foam surface used in this paper. 
This foam had 80 fibers of radius 0.01 in a height of 1, giving it $A=10, D=4.3\%, \bar{u}=2.2$
This is a dense foam.}
\label{fig:geom}
\end{figure}

Foam was implemented as a collection of whiskers. The whiskers within one simulation 
 all had the same 
radii. Whisker radius, height of the simulation volume, and number of whiskers were 
the free parameters of the simulation. Random whiskers were placed uniformly distributed in 
space and solid angle. The whiskers were as long as fit within the simulation volume.
 An example of such a surface 
is depicted in Figure \ref{fig:geom}.

In comercially available foams, 
foam whiskers extend a finite distance rather than as long as fits within the foam layer. 
This is different from our Monte Carlo calculations.
 In Section \ref{sec:analytic}, we find 
that the SEY from a foam surface depends only on local parameters. Because 
of this, we expect our calculations to be applicable to foams with finite whisker 
length.

\section{Analytic model}

\label{sec:analytic}

To support the numerical results, we formulated an analytic model of secondary 
electron emission from foam. This analytic model is an extension of a model published in our 
previous paper \cite{swanson2016}. While our previous model 
considered a field of uniformly aligned whiskers ($\hat{a}=\hat{z}$), 
we consider a field of randomly aligned whiskers ($\hat{a}$ isotropic). 
Here, $\hat{a}$ is the direction of the whisker axis.

As in the velvet model, we consider only one generation 
of secondary electrons. No tertiary electrons will be 
considered.

As in the velvet model, we will assume that electrons 
inside a whisker layerl hit the whiskers 
with uniform probability per unit distance traveled 
perpendicular to the whiskers' axes. If the whiskers have 
radius $r$ and areal density (whiskers per unit area, where 
area is defined perpendicular to the axis) $n$, the 
probability of intersection with a whisker is

\begin{equation}
P(hit)=2rndS_\perp
\end{equation}

where $dS_\perp$ is the distance traveled perpendicular 
to the axis. If the whiskers are aligned along $\hat{a}$, this becomes 

\begin{equation}
P(hit)=2rn\frac{\sqrt{1-\hat{v}\cdot\hat{a}^2}}{\hat{v}\cdot\hat{z}}dz
\end{equation}

where $z$ is the direction normal to the solid surface and $\hat{v}$ is the 
direction of primary electron incidence. This equation is linear with density 
$n$ of whiskers. Different populations of whiskers $1,2$ will add:

\begin{equation}
P_{1+2}(hit)=2rn_1\frac{\sqrt{1-\hat{v}\cdot\hat{a_1}^2}}{\hat{v}\cdot\hat{z}}dz+2rn_2\frac{\sqrt{1-\hat{v}\cdot\hat{a_2}^2}}{\hat{v}\cdot\hat{z}}dz
\end{equation}

Since $\hat{a}$ is isotropic, $l\equiv\hat{a}\cdot\hat{v}$ is uniformly 
distributed. Thus in a field of infinitely many 
infinitesimally dense fields of isotropically aligned 
whiskers, the probability of hitting one is 

\begin{equation}
P(hit)=2rn \frac{dz}{\mu} \int^{1}_{-1}dl \frac{\sqrt{1-l^2}}{2} = \frac{\pi}{2} rn\frac{dz}{\mu}
\end{equation}

where $\mu\equiv \hat{v}\cdot\hat{z}$. 

The probability that an electron will traverse $\Delta z$ 
without having hit a whisker is this value integrated over 
$z$

\begin{equation}
\label{eqdeltaz}
P(\Delta z)=e^{-\frac{\pi}{2}rn \frac{\Delta z}{\mu}} = e^{\frac{\bar{u}}{\mu}\frac{\Delta z}{h}}
\end{equation}

We have discovered the important parameter to describe 
the SEY reduction from a foam, $\bar{u}\equiv \frac{\pi}{2} rhn = AD/2$ 
where $A$ is the aspect ratio $A\equiv h/r$ and $D$ is the volume fill density. $\bar{u}$ is a measure of how 
much whisker there is: the more dense, or long, or wide the 
whiskers, the higher $\bar{u}$. It is related ($\bar{u}=\frac{\pi}{4}u$) to the 
parameter $u$ found for velvet, with the differences in geometry 
accounting for the numerical coefficient \cite{swanson2016}. 

The reduction from SEY can be interpreted as the probability 
that a secondary electron will escape from the whisker layer

\begin{equation}
\gamma_{eff}=\gamma P(escape).
\end{equation}

The electrons may be produced either at the top of the foam, at the bottom surface 
where the foam meets the solid substrate, or on the sides 
of the whiskers,

\begin{equation}
\gamma_{eff}=\gamma_{top}+\gamma_{bottom}+\gamma_{sides}.
\end{equation}

We shall now determine the value of each of these.

If a primary electron hits the top of the foam region, where the
 foam meets the vacuum, all secondary electrons will 
be freely conducted to the bulk. A primary electron hits the top 
with probability $D$, as this is the proportion of the 
top surface which is taken up by material.

\begin{equation}
\gamma_{top}=D\gamma
\end{equation}

The SEY from the bottom surface is:

\begin{equation}
\gamma_{bottom}=\gamma P(escape|bottom) P(bottom)
\end{equation}

The probability that a primary electron will make it to 
the bottom surface is derivable from equation \ref{eqdeltaz}. 

\begin{equation}
P(bottom)=e^{-\frac{\bar{u}}{\mu}}
\end{equation}

The probability that a secondary electron will escape after being
 emitted from the bottom depends on its emitted polar angle and in integrated form is 

\begin{equation}
P(escape|bottom)=\int^1_0 d\mu_2 P(escape|\mu_2,bottom) P(\mu_2)
\end{equation}

where $P(\mu_2)d\mu_2$ is the probability density function (PDF)
of $\mu_2=\cos\theta_2$, the polar angle of the secondary 
electron. Assuming a cosine distribution for the probability 
of polar angles of secondary electrons \cite{Bronstein}, $P(\mu_2)=2\mu_2$. 

$P(escape|\mu_2,bottom)$ is also calculable from Equation 
\ref{eqdeltaz}, yielding a final bottom SEY of

\begin{equation}
\gamma_{bottom}=2\gamma(1-D) \int^1_0 d\mu_2 \mu_2 e^{-(\frac{1}{\mu}+\frac{1}{\mu_2})\bar{u}}
\end{equation}

The procedure is similar for $\gamma_{sides}$, except that 
the secondary electron may be emitted at any $z$ value from 
0 to $h$. 

\begin{equation}
\label{eq:firstsides}
\gamma_{sides}=\left\langle\gamma\right\rangle  (1-D) \int^h_0 dz P(escape|z)P(z)
\end{equation}

Again $P(z)dz$, the PDF that an electron hit within $dz$ may 
be derived from Equation \ref{eqdeltaz}.

$\left\langle\gamma\right\rangle$ is necessary as, according to the 
empirical model of Vaughan \cite{vaughan}, SEY from a primary 
electron which is shallowly incident is larger than SEY from a 
primary electron which is normally incident. According to Equation
 \ref{eq:angularsey} and the value of $k_s=1$ of a smooth
 surface, the average SEY from isotropically aligned surface 
 elements will be larger than that of the flat value by
 
 \begin{equation}
 \label{eq:avgsey}
 \left\langle\gamma\right\rangle/\gamma=\int^1_0 d(\cos\theta) 2\cos\theta (1+\frac{\theta^2}{2\pi}) \approx 1.12
 \end{equation}
 
 Keeping explicit 
the dependence on $\mu_2$, Equation \ref{eq:firstsides} may be written

\begin{equation}
\gamma_{sides}= \left\langle\gamma\right\rangle(1-D) \int^1_0 d\frac{z}{h} \int^1_0 d\mu_2 \frac{\bar{u}}{\mu}e^{-\bar{u}\frac{z}{h}(\frac{1}{\mu}+\frac{1}{\mu_2})}P(\mu_2|\mu)
\end{equation}

Carrying out the $z$ integration

\begin{equation}
\gamma_{sides}= \left\langle\gamma\right\rangle(1-D) \int^1_0 d\mu_2 \frac{1-e^{-\bar{u}(\frac{1}{\mu}+\frac{1}{\mu_2})}}{1+\frac{\mu}{\mu_2}}P(\mu_2|\mu)
\end{equation}

The function $P(\mu_2|\mu)$ is the probability that a primary 
electron with polar velocity vector component $\mu=\cos\theta$ 
will produce a secondary electron with polar velocity 
component $\mu_2=\cos\theta_2$. Clearly this depends on where 
on a fiber this electron hits, and how the fiber is aligned.

Here we appeal to geometrical reasoning. Since $\hat{a}$, the 
whisker axes, are isotropically distributed, so too are 
$\hat{n}$, the vectors normal to the surface elements on the 
sides of the whiskers. Because of this, the probability 
that a primary electron hits a surface element with normal 
$\hat{n}$ will be linearly weighted by $\hat{v}\cdot\hat{n}$.

Integrating over all surface element normal vectors, 

\begin{equation}
\label{eq:pmu2mu1}
P(\mu_2|\mu)=\frac{4}{\pi} \int^1_{-1} dm (A_1\sin\phi_1+B_1\phi_1)(A_2\sin\phi_2+B_2\phi_2)
\end{equation}
\begin{equation*}
A_{1,2}=\sqrt{(1-m^2)(1-\mu_{1,2}^2)},B_{1,2}=m\mu_{1,2}
\end{equation*}
\begin{equation*}
 \phi_{1,2}=Re[\cos^{-1}(-\frac{B_{1,2}}{A_{1,2}})]
\end{equation*}

where $Re(x)$ is the real part of $x$. $m$ is an integration variable, but it may be informative 
to know that $m=\hat{n}\cdot\hat{z}$.

Thus the total SEY from a foam surface is expected to be

\begin{equation}
\label{eq:finalsey}
\begin{split}
\gamma_{eff}=\gamma D + (1-D)[\gamma \int^1_0 d\mu_2 2\mu_2 e^{-(\frac{1}{\mu}+\frac{1}{\mu_2})\bar{u}} \\+ \left\langle\gamma\right\rangle \int^1_0 d\mu_2 \frac{1-e^{-\bar{u}(\frac{1}{\mu}+\frac{1}{\mu_2})}}{1+\frac{\mu}{\mu_2}}P(\mu_2|\mu)]
\end{split}
\end{equation}

where $\left\langle\gamma\right\rangle$ is defined in Equation \ref{eq:avgsey} and $P(\mu_2|\mu)$ is defined 
in Equation \ref{eq:pmu2mu1}. Recall that $\mu=\cos\theta$, where $\theta$ is the polar angle. 
Also recall that $D$ is the volumetric fill ratio and $\bar{u}=AD/2$ where $A=h/r$ the ratio 
between the whisker layer thickness and the whisker radius.

\begin{figure}[tbp]
\centering\includegraphics[scale=0.60]{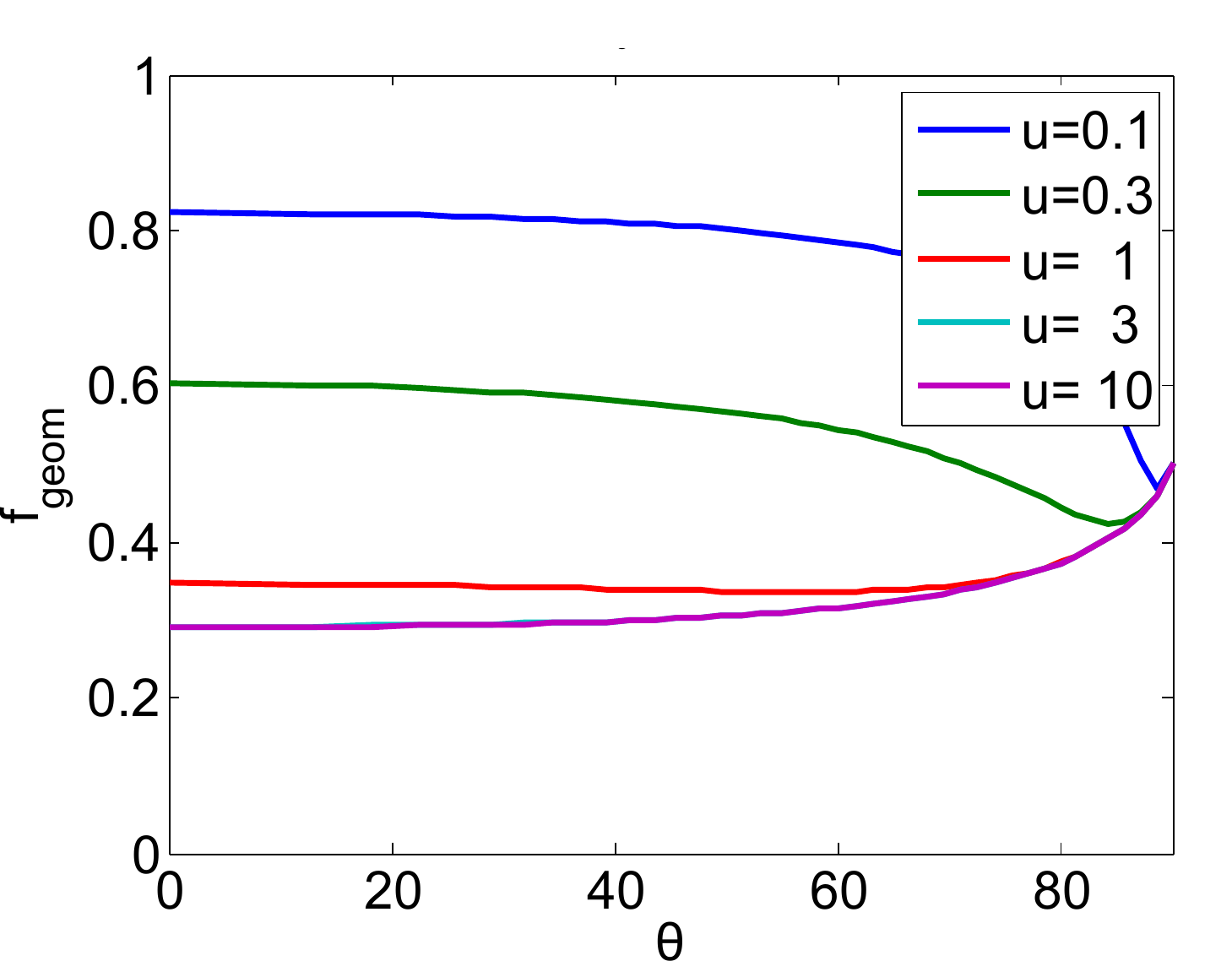}
\caption{Results of analytic theory. Total SEY is 
$\gamma[D+(1-D)f_{geom}]$.}
\label{fig:analytic}
\end{figure}

The factor in the square brackets is a function only of $\bar{u}$ and $\theta$. It is plotted in 
Figure \ref{fig:analytic}. 

For the case of isolated hard-sphere balls of radius $r$, volume density 
$n$, and layer height $h$, the analytical calculation for SEY is identical.
This includes the value of $P(\mu_2|\mu)$. For this 
case, 

\begin{equation}
\bar{u}_{ball}=\pi r^2 nh
\end{equation}

\section{Results and explanantion}

\label{sec:results}

\begin{figure}[tbp]
\centering\includegraphics[scale=0.60]{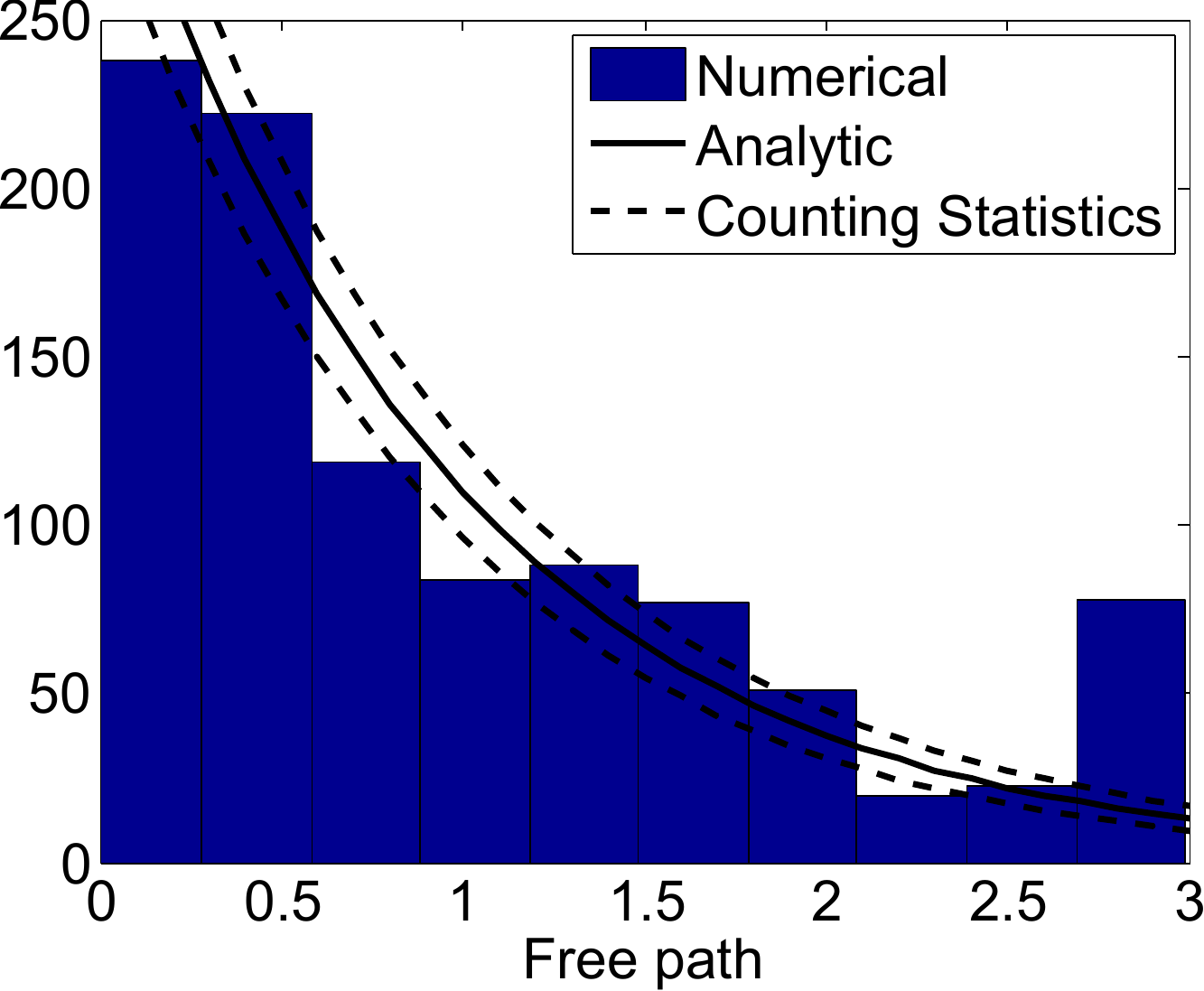}
\caption{Mean Free Path comparison: A histogram of free paths calculated 
during a Monte Carlo simulation compared with the idealized analytic version. 
Error arises from the counting statistics of both particles and whiskers.}
\label{fig:mfp}
\end{figure}

\begin{figure}[tbp]
\centering\includegraphics[scale=0.60]{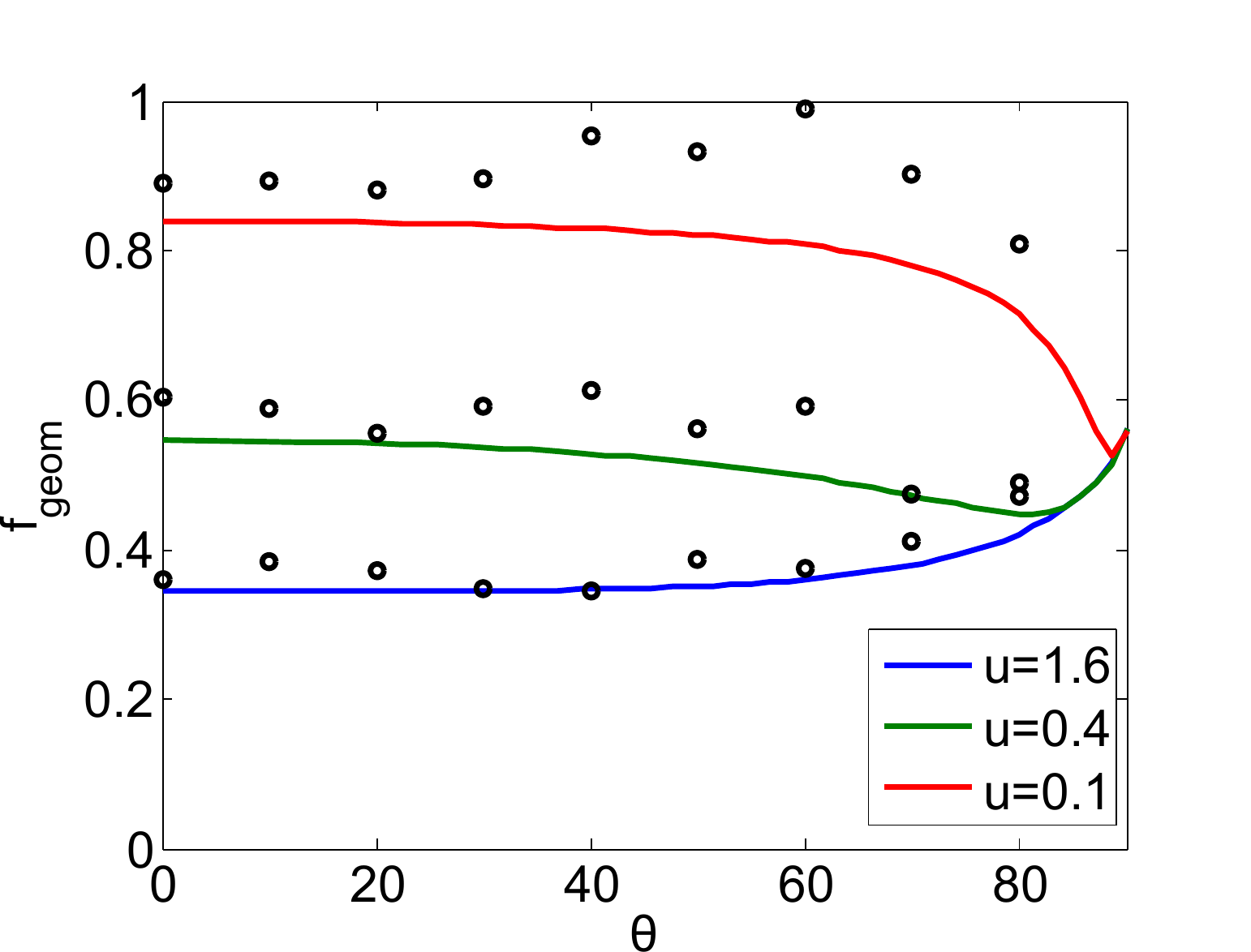}
\caption{Results of analytic theory compared to 
full numerical Monte Carlo model.}
\label{fig:numerical}
\end{figure}

\begin{figure}[tbp]
\centering\includegraphics[scale=0.60]{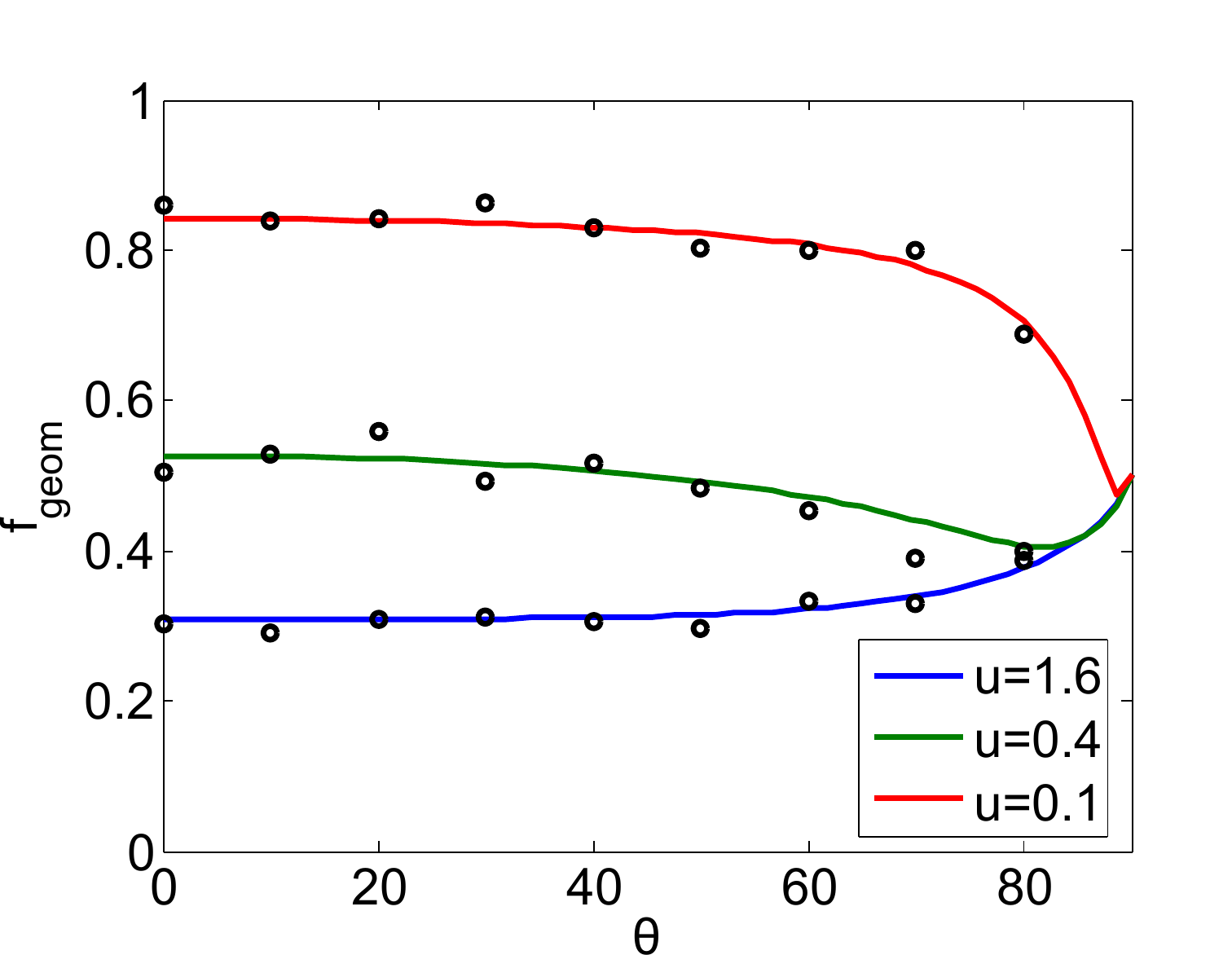}
\caption{Results of analytic theory compared to simplified Monte Carlo 
model consisting of only one generation of secondary electrons and 
no dependence of SEY on primary angle of incidence. }
\label{fig:diagnostic}
\end{figure}

The analytic model is based on the assumption that the mean free path is

 \begin{equation}
 \lambda_{mfp}=(\frac{\pi}{2}rn)^{-1}
\end{equation}

 To verify this, we tabulated the free paths of electrons within the foam layer during a 
 Monte Carlo calculation.

The results are plotted in histogram form in Figure \ref{fig:mfp}. For this calculation, 
whisker parameters were $r=0.005, h=3$, and 160 whiskers were in the simulation 
volume. This produced a $\bar{u}=3.2$ and a $\lambda_{mfp}=0.94$. The figure 
indicates that the assumption is qualitatively justified. The excess at a free path 
of 3 is the result of electrons hitting the bottom surface.

The normalized SEY as a function of primary angle of incidence and the 
$\bar{u}$ factor is plotted in Figure \ref{fig:numerical}. Three values of $\bar{u}$ are 
plotted: The $\bar{u}=0.1$ run was initialized with whisker layer height $h=3$, 
whisker radius $r=0.0025$, and 10 whiskers total in this volume. The $\bar{u}=0.4$ run was initialized with whisker layer height $h=3$, 
whisker radius $r=0.0025$, and 40 whiskers total in this volume. The  $\bar{u}=1.6$
 run was initialized with whisker layer height $h=3$, 
whisker radius $r=0.005$, and 80 whiskers total in this volume. 

We can see from 
Figure \ref{fig:numerical} that the analytic theory consistently under-estimates the 
SEY from a given foam by about $10\%$. The source of this discrepancy is subsequent generations of secondary electrons. 
In the analytic model, only one generation of secondary electrons is considered. 
In in Figure \ref{fig:diagnostic}, tertiary electrons are disabled. The numerically 
and analytically calculated results in Figure \ref{fig:diagnostic} are consistent.

The behavior of $\gamma$ at very small $\bar{u}$ can be explained thusly:
When there are very few whiskers, or they are very thin, or the whisker layer 
is very short, the probability of interacting with a whisker is small and so 
SEY is not reduced by much. 

The behavior of $\gamma$ at shallow angles of incidence ($\theta \rightarrow 90^\circ$) 
can be explained simply. A primary electron that is shallowly incident will 
hit a whisker very close to the top of the whisker layer. Because of isotropy 
of the whisker axes, this electron will have a 0.5 probability of being emitted 
with velocity in the upward hemisphere and a 0.5 probability of being emitted
with velocty in the downward hemisphere. Thus the SEY from shallow incidence 
will be reduced by one-half.

The behavior at more normal angles (low $\theta$) at high $\bar{u}$ is very isotropic.
There is very little dependence on the angle. This is expected: As $\bar{u}$ 
increases, almost no electrons penetrate to the bottom surface. If the bottom surface 
is not relevant, the problem is entirely isotropic. Interestingly, the value of minimum 
SEY reduction factor is close to $1/e$, the inverse Euler's constant. This is the value 
expected from an isotropically scattering and absorbing medium, such as a field of
infinitesimal hard spheres. 

\section{Conclusions}

We calcuated the SEY from a foam surface and verified that it is reduced. Furthermore 
our calculations support the prediction that SEY from a foam surface will behave more isotropically 
than from other fiberous surfaces like velvet. We find that foam cannot reduce SEY by more than 
about $30\%$ of its un-suppressed value. We find that foam does not suppress SEY as much 
as velvet given the same geometric factors $A,D$. 

\subsection{Acknowledgment}

The authors would like to thank Yevgeny Raitses, who attracted our attention
to the SEY of complex surfaces. We would also like to thank Eugene Evans, 
who suggested the approach of isosurfaces. This work was conducted under a subcontract with
UCLA with support of AFOSR under grant FA9550-11-1-0282.

\end{document}